\documentclass[11pt]{article}

\usepackage{amsmath,amssymb}
\usepackage{graphicx}
\newcommand{\be}{\begin{equation}}
\newcommand{\ee}{\end{equation}}
\newcommand{\bea}{\begin{eqnarray}}
\newcommand{\eea}{\end{eqnarray}}
\newcommand{\ba}{\begin{array}}
\newcommand{\ea}{\end{array}}

\newcommand{\alp}{\alpha'}

\newcommand{\tr}{\mbox{tr}}

\newcommand{\p}{\partial}

\newcommand{\la}{\langle}
\newcommand{\ra}{\rangle}

\newcommand{\cP}{{\cal{P}}}

\begin{document}

\begin{titlepage}
\begin{flushright}
KEK-TH-1211
\end{flushright}

\vfill

\begin{center}
{\Large \bf
Production cross section of rotating string
}

\bigskip\bigskip\bigskip

{\large Tsunehide Kuroki\footnote{\tt tkuroki@post.kek.jp} 
and
Toshihiro Matsuo\footnote{\tt tmatsuo@home.phy.ntnu.edu.tw}
}

\bigskip\bigskip

{\it
$^1$Theory Group, KEK, Tsukuba, Ibaraki 305-0801, Japan\smallskip\\
$^2$Department of Physics,
National Taiwan Normal University,
Taipei City 116, Taiwan
}

\end{center}

\vfill
\begin{abstract}
We calculate production cross sections of a single rotating string from a collision of two light states in bosonic string theory.
We find that the cross sections are written in terms of the modified Bessel function of the first kind with the degree given by the angular momentum in the high energy regime.
We also obtain a similar formula from the partial wave expansion of four point amplitudes. Asymptotic behavior of the cross sections is consistent with a well known form factor of a string. 

\end{abstract}
\vfill

\end{titlepage}

\setcounter{footnote}{0}
\section{Introduction}
\label{intro}
\setcounter{equation}{0}

Ultra-high energy scattering of strings is a genuine probe 
into their underlying fundamental structure which is invisible 
in ordinary circumstances.
Many interesting features have emerged such as among other things higher genus dominance in hard processes \cite{Gross:1987kz, Gross:1987ar}, 
saturation on the Cerulus-Martin lower bound in saddle point analysis \cite{Mende:1989wt}, partial wave unitarity \cite{Soldate:1986mk, Muzinich:1987in}, linear relations \cite{Gross:1988ue, Chan:2003ee, Chan:2005zp}, an appearance of the shock wave metric in the eikonal amplitudes in Regge processes \cite{Amati:1987wq, Amati:1987uf, Amati:1988tn, Veneziano:2004er}\footnote{See \cite{Amati:2007ak} for recent progress.}, and spacetime uncertainty \cite{Yoneya:2000bt}, etc.
Still recently the perturbative high energy scattering amplitudes have been attracted lots of attention and been used to study for various purposes especially in the new light of AdS/CFT \cite{Polchinski:2001tt, Polchinski:2002jw, Brower:2006ea, Alday:2007hr}, etc.
In the investigations mentioned above basic ingredients are the four point elastic scattering amplitudes of the Veneziano and the Virasoro-Shapiro.
On the other hand, three point amplitudes, which are just constants 
and less interesting when the external states are in lower level 
tachyons or massless states, become nontrivial when at least one of the external states is set in an extremely highly excited state.

However in general such a highly excited state has a complicated expression of vertex operator.
In \cite{Amati:1999fv} Amati and Russo have studied such a three point process including highly excited states and indeed they have shown that a single highly exited string emits a light state and behaves 
as a black body with the Hagedorn temperature which is the characteristic temperature of string theory, 
when the initial states are averaged and the final states are summed over the degeneracy which are in the same mass level.
They calculated directly a decay rate instead of an amplitude 
in such a way that is quite similar to the unitarity relation 
(optical theorem) by introducing a projection operator 
to realize highly excited string states as external states 
which in general are difficult to treat in perturbative calculations. 
Their idea has been extended and applied to subsequent works in {\it e.g.}  \cite{Manes:2001cs, Manes:2003mw, Manes:2004nd, Chen:2005ra}.

Another interesting application of the three point amplitude is to pursue processes of black hole formation in ultra-high energy scatterings and issues of unitarity both of which are closely related. 
In \cite{Dimopoulos:2001qe} Dimopoulos and Emparan have investigated, among other things, production of a single highly excited string as a black hole progenitor, in view of the correspondence principle \cite{Horowitz:1996nw,Horowitz:1997jc,Damour:1999aw}. 
They conclude that the total production cross section of a highly excited string is rather large and comparable to the conjectured black hole geometric cross section, though these two do not match 
at the correspondence point.
The cross section of a string grows much faster than that of black hole 
and thus seems to violate the unitarity bound.

All of these arguments so far have been made without specifying 
the angular momentum or in other words the impact parameter 
in the case of scattering events.
Namely, these amplitudes are obtained with summing over all contributions from states which have different angular momenta. 
It is important to calculate more detailed amplitudes or cross sections with the angular momentum specified because the cross sections or S-matrices are basic data that characterize the nature of fundamental strings. 
Furthermore they are also useful 
for other practical purposes to reveal various features of a string. 
For instance, specifying angular momentum or impact parameter provides 
a starting point to argue non-locality \cite{Giddings:2002cd, Giddings:2006vu, Giddings:2006sj, Giddings:2007bw, Giddings:2007qq}, and 
decay rates of a specific state which has maximal angular momentum have been investigated in \cite{Iengo:2002tf, Iengo:2003ct, Chialva:2003hg, Chialva:2004xm, Iengo:2006gm}.

In this paper we shall calculate a production cross section 
of a single rotating string from a collision of two strings, 
which is described by a three point amplitude via two methods.
The first one is to introduce a projection operator which specifies eigenstates of a definite component of the angular momentum, say, 
$J_{12}$ as done in \cite{Amati:1999fv} for the mass level.
With a help of this operator we obtain a closed formula 
for the production cross section of a single string 
with a definite angular momentum. 
Our formula agrees with \cite{Dimopoulos:2001qe} when we sum 
over the angular momentum. 
A similar result can be obtained via the optical theorem 
for a four point amplitude where the total angular momentum 
of a single produced string is specified.


\smallskip

This paper is organized as follows.
In section two we calculate production cross sections 
of a heavy string produced from a collision of two light states 
without specifying the angular momentum. 
We employ both tachyon-tachyon and photon-photon collisions. 
In section three we calculate a production cross section 
of a process in which we specify one of components 
of the angular momentum.
In section four we derive from the partial wave expansion 
of a four point amplitude a production cross section 
of a rotating string with the total angular momentum specified. 
In appendices we present details of formulae used in our calculations.

\section{Collision between two light states}
\label{withoutJ}
\setcounter{equation}{0}
In this section we calculate production cross sections 
of a single heavy string from a collision between two light states 
in the bosonic (oriented) open string theory.
Here the collision is inclusive with respect to the angular momentum, 
that is, we do not specify the angular momentum of the process. 
Namely we sum over all the angular momenta.

\subsection{tachyon-tachyon collision }
As a warming-up exercise, we consider a tachyon-tachyon collision.
Let the initial tachyon momenta be $k_\mu$ and $p_\mu$, and the momentum of a final massive string state be $P_\mu$, 
which is assumed to be in a level $N$ state.
The final state is denoted by $\Phi_{N}$. 
The on-shell conditions for the initial state tachyons are
\bea
k^2=p^2=1,
\eea
and for the final state
\bea
P^2=-N+1, 
\eea
where we set $\alp=1$.
Let $s=-(p+k)^2$, then from the energy-momentum conservation $P_\mu=p_\mu+k_\mu$ (on-shell condition which will see later) we have
\bea
s=N-1.
\eea

We consider a probability of the collision process which is given as
\bea
\cP(V(p),V(k) \to \Phi_{N})
=\sum_{\Phi|N}|\la \Phi|
\frac{1}{\sqrt{2}}\left(V(k)+\Theta V(k)\Theta^{-1}\right)
|p\ra|^2,
\eea
where the symbol $\Phi|N$ means that we sum all over the states $\Phi$ in the level $N$.
This is because we do not observe detailed information 
on the final state.
They are specified only by the level and thus will have lots of degeneracy as they go to a higher level.
Here the tachyon vertex operator\footnote{
Here the argument $z=1$ in $X(z)$ denotes the origin in the coordinate system $(\tau,\sigma)$ which is related to $z$ as $z=e^{\tau+i\sigma}$.}
is
\bea
V(k,1)=g_o : \exp(i k \cdot X(1)) : ,
\eea
where $g_o$ is the open string coupling constant.
And $\Theta=(-1)^{\hat{N}}$ is the twist operator which takes 
a vertex operator on one boundary to the other boundary.
Here $\hat{N}=\sum_{n \mu} \alpha_{-n \mu} \alpha_n^\mu$ is the number operator
and $\alpha_n$ obeys the commutation relation 
$[\alpha_n, \alpha_m]=n \delta_{n-m}$.

We introduce an operator which is a projection 
to the level $n$ state \cite{Amati:1999fv}:
\bea
P_n = \oint \frac{dz}{2\pi i z} z^{\hat{N}-n} ,
\eea 
which enables us to write the probability as
\bea
\cP(V(p),V(k) \to \Phi_{N})
&=& 
\frac{1}{2}\sum_{i,j=all}
(
\la i|V(-k)P_N|j\ra\la j|V(k)P_0|i\ra
+\la i|\Theta V(-k)\Theta^{-1} P_N|j\ra\la j|V(k)P_0|i\ra
\nonumber \\
&&\hspace{-1cm}+
\la i|V(-k)P_N|j\ra\la j|\Theta V(k)\Theta^{-1}P_0|i\ra
+\la i|\Theta V(-k)\Theta^{-1} P_N|j\ra\la j|\Theta V(k)\Theta^{-1}P_0|i\ra
)
\nonumber \\
&=& \tr[
V(-k,1)P_NV(k,1)P_0+V(-k,1)P_N V(k,-1) P_0
],
\eea
where $i,j$ take all the physical states in Hilbert space and we have used $P_N=\Theta P_N \Theta^{-1}$. 
Here the trace is taken only on the oscillator parts.
As for the zero mode, it is easy to evaluate
\bea
|\la P| e^{i k\cdot \hat{x}} |p\ra|^2
=V\delta(P-p-k).
\label{onshell}
\eea
Hereafter we will omit this factor for simplicity and the on-shell condition should be understood.
The position of the vertex operator is converted as
\bea
z^{\hat{N}}V(k,1)z^{-\hat{N}}=V(k,z) , 
\eea
thus we can write the probability as follows
\bea
\cP(V(p),V(k) \to \Phi_{s+1})=
\oint_{C_{z_1}} \frac{dz_1}{2\pi i z_1}z_1^{-s-1}
\oint_{C_{z_2}} \frac{dz_2}{2\pi i z_2}
\tr[V(-k,1)(V(k,z_1)+V(k,-z_1))(z_1z_2)^{\hat{N}}],\nonumber\\
\eea
where the contours $C_{z_1},C_{z_2}$ should be taken as $|z_1|,|z_2|<1$ so that they will not hit the vertex operators.
By change of variables $z_1=v, z_1z_2=w$ (thus $|w|<|v|<1$) we have
\bea
\cP(V(p),V(k) \to \Phi_{s+1})=
\oint \frac{dv}{2\pi i v}v^{-s-1}
\oint \frac{dw}{2\pi i w}
\tr[V(-k,1)(V(k,v)+V(k,-v))w^{\hat{N}}] .
\eea
Note that there is no momentum integration.  
This can be seen as a sum of two (oscillator part of) one-loop cylinder amplitudes, one with two vertex operators inserted at the same boundary (planar diagram) and the other with those inserted at both boundaries (non-planar diagram).
The trace part is evaluated as 
(see appendix \ref{tracecalc} for detail) 
\bea
\tr[V(-k,1)V(k,v)w^{\hat{N}}] =
[f(w)]^{2-D} \left[\hat{\psi}(v,w)
\right]^{-2}, 
\label{traceresult0}
\eea
where $D$ is the number of spacetime dimensions and $f(w)$ and $\hat{\psi}(v,w)$ are defined in (\ref{Dedekind}) and (\ref{psihat}), respectively.

Now it is straightforward to perform the $w$ contour integration because it only picks up constant terms in the expansion of the integrand with respect to $w$.
One obtains
\bea
\cP(V(p), V(k) \to \Phi_{s+1})&=&g_o^2 \oint \frac{dv}{2\pi i v}v^{-s-1}
\left[\frac{1}{(1-v)^2}+\frac{1}{(1+v)^2}\right]
\nonumber \\
&=&g_o^2(s+2)(1+(-1)^{s+1}) .
\label{tachyonprob}
\eea
Dividing phase space factor we get a cross section:
\bea
\sigma_{\text{open}} \simeq g_o^2 (1+(-1)^{s+1}),
\eea
where we have taken large $s$. 

In the case of a closed string collision, calculation may be done 
in a similar way as in the case of the open string.
Actually, at the tree level, a squared amplitude of open strings provides that of closed strings \cite{Kawai:1985xq}.
Both planar and non-planar diagrams give rise to the same closed string amplitude.
Therefore for a collision with two closed string tachyons we have a probability: 
\bea
\cP(V(p), V(k) \to \Phi_{s+1})_\text{closed} \sim g_s^2 s^2 ,
\eea
or cross section:
\bea
\sigma_\text{closed} \simeq g_s^2 s ,
\eea
which is precisely the one derived in \cite{Dimopoulos:2001qe}.

\subsection{photon-photon collision}
At high energy a cross section should not depend on the type 
of initial states.
To check this we calculate a cross section of an inclusive process 
of a collision of two photons.
Actually we will obtain a cross section which is essentially the same 
as that of the tachyon-tachyon case at high energy.
 
We consider initial photons with polarization and momenta 
$(\zeta^\mu, k_\mu)$ and $(\xi^{\mu}, p_\mu)$ 
and a final massive string state $\Phi$ of level $N=s+1$.
The initial state polarization is averaged and that of the final state is summed.
The probability is
\bea
\cP(\xi(p), \zeta(k) \to \Phi_N)
=\frac{1}{(D-2)^2}\sum_{\xi,\zeta}\sum_{\Phi|N}|\la \Phi|V(\zeta,k)|\xi,p\ra|^2.
\eea
We will concentrate on the planar contribution 
until the last step of the computation. Contribution 
from the non-planar diagram is easy to include once we obtain 
the planar part as we saw in the last subsection.
Here the photon vertex operator is 
\bea
V(\zeta,k,1)=g_o \exp(i k \cdot X(1)+\zeta \cdot \dot{X}(1))|
_{\text{linear\,\, in}\,\, \zeta},
\label{photon_vertex_op}
\eea
where the normal ordering is irrelevant 
because $k^2=k \cdot \zeta=0$ and only the linear term in $\zeta$ 
is relevant for us.

With the help of the projection operator 
we have
\bea
\cP(\xi(p), \zeta(k) \to \Phi_N)=\frac{1}{(D-2)^2}\sum_{\zeta}
\oint \frac{dw}{2\pi i w}w^{-1}
\oint \frac{dv}{2\pi i v}v^{1-N}
\tr[V(\zeta,-k,1)V(\zeta,k,v)w^{\hat{N}}].
\eea
The trace part can be calculated as
\bea
\tr[V(\zeta,-k,1)V(\zeta,k,v)w^{\hat{N}}]
=
g_o^2 f(w)^{2-D} \zeta^2 \hat{\Omega}(v,w),
\label{photontrace}
\eea
where
\bea
\hat{\Omega}(v,w) 
=\sum_{n=1}^\infty n
\left(v^n+\frac{w^n(v^n+v^{-n})}{1-w^n}\right) ,
\eea
and we have
\bea
\cP(\xi(p), \zeta(k) \to \Phi_{s+1})=\frac{g_o^2}{D-2} \oint \frac{dw}{2\pi i w}w^{-1} 
f(w)^{2-D}
\oint \frac{dv}{2\pi i v}v^{-s} \hat{\Omega}(v,w),
\eea
where we have used $\sum_{\zeta} \zeta^2= (D-2)$ as well as the on-shell condition.
The integrations of $v$ and $w$ are easily performed and we obtain
\bea
\cP(\xi(p), \zeta(k) \to \Phi_{s+1})
=
g_o^2 s (1+(-1)^s) ,
\label{photonprob}
\eea
where we have added a contribution from the non-planar diagram which is obtained simply by replacing the argument $v$ to $-v$ in 
(\ref{photontrace}).
This is the same behavior as that of the previous two tachyons collision at large $s$.

We close this section by making a remark on the contribution 
from the non-planar diagram. 
In (\ref{tachyonprob}) or (\ref{photonprob}), the non-planar 
contribution exactly cancels the planar one 
when $s$ is even or odd, respectively. 
Since these probabilities 
are essentially the pole residues at $s=N-1$ of the corresponding 
four point amplitudes via the optical theorem 
as explained in section \ref{withL}, this observation reflects 
the fact that when all the four external lines are the same in the 
four point amplitude, its pole residues in the different channels 
have the same absolute value with different signs. 
Therefore, in a general case we do not expect such a cancellation. 
In fact, in contrast to our case, 
in \cite{Amati:1999fv} the non-planar contribution 
is shown to be subleading 
with respect to $N$ compared to the planar one.

\section{Collision with specifying angular momentum}
\label{withJ}
\setcounter{equation}{0}

Now we specify the angular momentum of a collision process.
We consider tachyons with momentum $k_\mu$ and $p_\mu$ to produce a massive string of level $N$ with an angular momentum.
To be precise, we take the angular momentum $J_{12}$.
The final state is denoted as $\Phi_{(N,J)}$.

The probability in which the final states are summed up 
over the degeneracy is
\bea
\cP(V(p),V(k) \to \Phi_{(N,J)})
=\sum_{\Phi|(N,J)}|\la \Phi|V(k)|p\ra|^2.
\eea
Again we will concentrate on the planar contribution until the last step of computation. 
 
Now we introduce an operator which is a projection to states with 
an angular momentum $J$:
\bea
Q_{J} = \oint \frac{dz}{2\pi i z} z^{\hat{J}-J} ,
\eea
where
\bea 
\hat{J}=\hat{x}^{1}\hat{p}^{2}-\hat{x}^{2}\hat{p}^{1}
-i \sum_{n=1}\frac{1}{n}
(\alpha_{-n}^{1} \alpha_n^{2} -\alpha_{-n}^{2}\alpha_n^{1}) .
\eea
Then the probability is written as
\bea
\cP(V(p),V(k) \to \Phi_{(N,J)})&=& 
\sum_{i,j=all}\la i|V(-k) Q_{J}P_N|j\ra\la j|V(k)P_0|i\ra
\nonumber \\
&=&\tr[V(-k,1) Q_{J}P_NV(k,1)P_0] .
\eea
Note that we do not need to insert $Q_{J=0}$ to pick up the spin zero tachyon state because this projection is already accomplished by $P_{N=0}$.
Here again the trace is taken only on the oscillator parts.
As for the zero mode, let us focus on the 1-2 components (the other components are the same as before):
\bea
\la p|v(-k)q_J|P\ra\la P|v(k)|p\ra
&=&\la p|e^{-i k \cdot \hat{x}}z^{\hat{x_1}\hat{p_2}-\hat{x_2}\hat{p_1}}|P\ra
\la P|e^{i k \cdot \hat{x}}|p\ra
\nonumber \\
&=&
\la p+k|z^{\hat{x_1}\hat{p_2}-\hat{x_2}\hat{p_1}}|P\ra
\delta(p+k-P) .
\eea
If we set the (1-2 component of) momentum of the final state zero, 
$P=0$, then
\bea
\la p+k|z^{\hat{x_1}\hat{p_2}-\hat{x_2}\hat{p_1}}|0\ra
=\delta(p+k),
\eea
and we obtain
\bea
V_2\delta^{(2)}(p+k) .
\eea
We will omit this factor in the following.
We use the on-shell condition to replace $N$ by $s+1$.

Then we have 
\bea
\cP(V(p),V(k) \to \Phi_{(s+1,J)})&=&
\oint_{C_{z}} \frac{dz}{2\pi i z}z^{-J}
\oint_{C_{u_1}} \frac{du_1}{2\pi i u_1}u_1^{-s-1}
\oint_{C_{u_2}} \frac{du_2}{2\pi i u_2}
\nonumber \\
&&\times
\tr[V(-k,1)z^{\hat{J}}V(k,u_1)(u_1u_2)^{\hat{N}}] .
\eea
The contours $C_{z},C_{u_1},C_{u_2}$ satisfy $|z|,|u_1|,|u_2|<1$ 
in such a way that they do not encounter the vertex operators.
Changing the variables as $u_1=v, u_1u_2=w$ (thus $|w|<|v|<1$) we get 
\bea
\cP(V(p),V(k) \to \Phi_{(s+1,J)})
=
\oint \frac{dz}{2\pi i z}z^{-J}
\oint \frac{dv}{2\pi i v}v^{-s-1}
\oint \frac{dw}{2\pi i w}
\tr[V(-k,1)z^{\hat{J}}V(k,v)w^{\hat{N}}] .
\eea
The trace part is computed in appendix \ref{tracecalc}.
Without loss of generality, one can set the $i=2, 3,..,d$ components of incoming momenta vanish.
Furthermore using the on-shell condition $k^2=1$, or $(k_1)^2=(k_0)^2+1$ which yields $s/4=(k_0)^2=(k_1)^2-1$, we have, including contributions from ghost,
\bea
\tr[V(-k,1)z^{\hat{J}}V(k,v)w^{\hat{N}}]
=
g_o^2
[f(w)]^{4-D}[\hat{\psi}(v,w)]^{s/2}
[\tilde{f}(z,w)]^{-1}\left[\hat{\Psi}(z,v,w)\right]^{-(s/4+1)} ,
\label{traceresult}
\eea
where $\tilde{f}(z, w)$ and $\hat{\Psi}(z,v,w)$ are defined 
in (\ref{tildef}) and (\ref{Psihat}), respectively.
It is evident that (\ref{traceresult}) reproduces (\ref{traceresult0}) 
when $z=1$, as expected. 
Again it is straightforward to perform the contour integrations with respect to the variables $w$ and $z$.
As for the $w$ contour integration only the $w=0$ term survives and the $z$ integration provides the modified Bessel function of the first kind\footnote{It can be defined by a contour integration:
\bea
I_\nu(t)=\oint \frac{dz}{2\pi i} z^{-\nu-1} e^{t/2(z+1/z)} .
\eea
}:
\bea
\oint \frac{dz}{2\pi i z}z^{-J} \oint \frac{dw}{2\pi i w}
\tr[V(-k,1) z^{\hat{J}}V(k,v)w^{\hat{N}}]
=
g_o^2 (1-v)^{s/2} I_{J}\left(-\left(\frac{s}{2}+2\right)\ln(1-v)\right) .
\eea
Thus we have
\bea
\cP(s,J)=g_o^2 \oint \frac{dv}{2\pi iv} v^{-s-1}(1-v)^{\tilde{s}/2-2} I_J\left(-\frac{\tilde{s}}{2} \ln(1-v)\right) ,
\label{expression1}
\eea
where $\tilde{s}:=s+4$.
An integral representation of the modified Bessel function\footnote{
That is
\bea
I_J(t)=\frac{(t/2)^J}{\sqrt{\pi}\Gamma(J+1/2)}\int_{-1}^1dz(1-z^2)^{J-1/2}e^{-tz}, \quad Re(J)>-1/2 .
\eea
}
enables us to write 
\bea
\cP(s,J)=\frac{g_o^2}{\sqrt{\pi}\Gamma(J+1/2)}
\int_{-1}^1 dz Q_{J-1/2}(z)
\oint \frac{dv}{2\pi iv} v^{-s-1}(1-v)^{\tilde{s}/2-2+z\tilde{s}/2}
\left(\frac{-\tilde{s}\ln(1-v)}{4}\right)^J ,
\label{exact}
\eea
where
\bea
Q_J(z)=(1-z^2)^{J} .
\label{Qfunction}
\eea
Then the $v$ integration can be easily done and we have 
\bea
\cP(s,J)=
\frac{g_o^2}{\sqrt{\pi}\Gamma(J+1/2)2^J}
\int_{-1}^{1} dz 
Q_{J-1/2}\left(z\right)
\left(-\frac{\p}{\p z}\right)^J 
\frac{\Gamma(s+3-(z+1)\tilde{s}/2)}{\Gamma(s+2)\Gamma(2-(z+1)\tilde{s}/2)} .
\label{exact z}
\eea
One can immediately show that the integrand is even or odd function of $z$ according as $s+J+1$ is even or odd integer, respectively.
Hence one can write
\bea
\cP(s,J)=
\frac{(1+(-1)^{s+J+1})g_o^2 (\tilde{s}/2)^{J-1}}{\sqrt{\pi}\Gamma(J+1/2)2^J}
\int_{0}^{\tilde{s}/2} dx 
Q_{J-1/2}\left(1-\frac{2x}{\tilde{s}}\right)
\left(-\frac{\p}{\p x}\right)^J 
\frac{\Gamma(s+3-x)}{\Gamma(s+2)\Gamma(2-x)} ,\nonumber\\
\label{exact1}
\eea
where we have changed the variable as 
\bea
x := (z+1)\tilde{s}/2 .
\eea

So far the expressions are exact. 
However, the integration in (\ref{expression1}) or (\ref{exact1}) 
may not be performed rigorously and hence we make an approximation.
Let us examine 
\bea
F(x):=
\frac{\Gamma(s+3-x)}{\Gamma(s+2)\Gamma(2-x)}
=
\frac{1}{\Gamma(s+2)}\left[(s+2-x)(s+1-x) \cdots(3-x)(2-x)\right],
\eea
in which the bracket part is so called the Pochhammer symbol. 
It is sufficient to consider $F(x)$ only in the integration region 
$0<x<\tilde{s}/2$. 
The first zero point is $x=2$ and in between $2<x<\tilde{s}/2$ 
it oscillates with taking quite small values of ${\cal O}(1/s^2)$ 
compared to those in $0<x<2$. 
Thus the integration region 
can be restricted to $0<x<2$, where we can neglect 
$\Gamma (2-x)$ since it gives only an ${\cal O}(1)$ contribution. 
For large $s$ 
we may use the Stirling formula to get
\bea
F(x) \sim s^{1-x} . 
\label{Stirlingapprox}
\eea
This is the only approximation we made in the calculation, 
which becomes better as $s$ becomes larger. 
Therefore we have
\bea
\left(-\frac{\p}{\p x}\right)^J
F(x)
\sim
(\ln s)^J s^{1-x},
\label{Stirling}
\eea
and (\ref{exact1}) becomes
\bea
\cP(s,J)
\simeq
\frac{(1+(-1)^{s+J+1})g_o^2 (\tilde{s}/2)^{J-1}}{\sqrt{\pi}\Gamma(J+1/2)2^J}
\int_{0}^{\tilde{s}/2} dx 
Q_{J-1/2}\left(1-\frac{2x}{\tilde{s}}\right)
(\ln s)^J s^{1-x} ,
\eea
where we recover the original integration region 
because the integrand is again sufficiently small in $2<x<\tilde{s}/2$. 
Thus we finally obtain the probability:
\bea
\cP(s,J)
\simeq
g_o^2 s^{1-\tilde{s}/2}I_J\left(\frac{\tilde{s}}{2}\ln s\right)
(1+(-1)^{s+J+1})(1+(-1)^{s+1}) ,
\label{J12crosssection}
\eea
where we have put the contribution from the non-planar diagram which is obtained by replacing the argument $v$ to $-v$ in (\ref{traceresult}).
It should be checked that summing over all the angular momenta reproduces the previous result (\ref{tachyonprob}) up to a term suppressed with $s$ (thus negligible at high energy):
\bea
\sum_{J=-\infty}^\infty \cP(s,J)=\cP(s),
\eea
where we have used $\sum_{J=-\infty}^\infty I_J(z)=e^z$ and $\sum_{J=-\infty}^\infty (-1)^JI_J(z)=e^{-z}$.

The cross section $\sigma(s,J)=\cP(s,J)/s$ is given as 
\bea
\sigma(s,J)
\simeq
g_o^2 e^{-\xi} I_J(\xi)(1+(-1)^{J})(1+(-1)^{s+1}) , \quad (\xi := \frac{\tilde{s}}{2}\ln s) .
\label{Jcrosssection}
\eea
This is one of the main results of this paper. 
The fact that this cross section vanishes for even $s$ 
may have the same reason as in (\ref{tachyonprob}). 
On the other hand, it should be noted that it also vanishes 
for odd $J$. We regard this result as a reflection 
of the fact that 
the system has a symmetry under the $\pi$ rotation 
of 1-2 plane $\varphi\rightarrow \varphi+\pi$ and that 
the wave function of the produced rotating string 
is rotationally symmetric in the 1-2 plane.

Although this expression is obtained by an approximation for large $s$, it is instructive to make a further approximation for the modified Bessel function with large $\xi$  to see the asymptotic behavior of the cross section. 
In appendix \ref{Ij} we give an asymptotic form 
of the modified Bessel function by a saddle point method. 
We find (up to the factor $(1+(-1)^{J})(1+(-1)^{s+1})$)
\bea
\sigma(s,J) \sim
g_o^2 \frac{\exp\left(-\xi+\xi\sqrt{1+\frac{J^2}{\xi^2}}-J \ln\left(\frac{J}{\xi}+\sqrt{1+\frac{J^2}{\xi^2}}\right)\right)}
{\sqrt{2\pi \xi\sqrt{1+\frac{J^2}{\xi^2}}}} .
\label{MBesselsaddle}
\eea
The leading term in the expansion of $J/\xi$ is Gaussian:
\bea
\sigma(s,J) 
\sim 
\frac{g_s}{\sqrt{2\pi \xi}}\exp\left(-\frac{J^2}{2\xi}\right), 
\label{JGaussian}
\eea
which explicitly shows that the cross section with a definite angular momentum decreases with energy, in particular 
$\sigma(s,J=0) \sim (s\ln s)^{-1/2}$.
The Gaussian factor tells us the form factor of a string.
By noting that the angular momentum is related to the impact parameter through $J=b\sqrt{s}/2$, the form factor corresponds 
to an extended object with size $\sqrt{\ln s}$ 
as is well known.

\section{Partial wave expansion}
\label{withL}
\setcounter{equation}{0}
In this section we give another derivation of a cross section 
with specifying the total angular momentum, or one component $J_{12}$ 
respectively, based on the optical theorem. 
In fact, by the optical theorem, we derive the total cross section 
as the imaginary part of the four-point amplitude. Then  
expanding it in terms of an appropriate base of eigenfunctions 
of the total angular momentum squared $\hat{L}^2$ or in terms of the one for $\hat{J}_{12}$, we can read off the cross section in question. 

We start with the Veneziano amplitude:
\bea
V(s,t,u)=A(s,t)+A(t,u)+A(u,s) ,
\eea
where
\bea
A(s,t)=g_{o}^2\frac{\Gamma(-1-s) \Gamma(-1-t)}{\Gamma(-2-s-t)} .
\eea
We make the partial wave expansion of an $s$-channel pole residue by using the Legendre polynomials which are eigenfunctions of the total angular momentum squared:
\bea
V(s,t,u)=\sum_L (2L+1) A_L(s) P_L(1+2t/\tilde{s}).
\eea
Note here that in the present case the scattering angle $\theta$ is related to the Mandelstam variables through 
$\cos\theta = 1+2t/\tilde{s}$.
We are interested in the imaginary part of the Veneziano amplitude. 
Noting
\bea
\text{Im}_{s \sim (N-1)}\Gamma(-s-1+i\epsilon) =\pi \frac{(-1)^{s+1}}{(s+1)!}\delta(s-N+1),
\eea
then one can easily see 
\bea
\text{Im}_{s \sim (N-1)} V(s,t,u)
=
g_{o}^2 \pi (1+(-1)^{s+1})
\frac{\Gamma(3+s+t)}{\Gamma(s+2)\Gamma(2+t)} \delta(s-N+1),
\eea
where the term proportional to $(-1)^{s+1}$ stems from $A(u,s)$.
We have other delta functions which represent the on-shell momentum conservation  condition for the pole residues. In the following we neglect the $\delta$-function factor 
with the on-shell condition understood. 
The coefficient function $A_L(s)$ in the partial wave expansion 
of this $s$-channel pole residue is
\bea
A_L(s)&=&\int_{-1}^1 dz \text{Im}V(s, t, u) P_L(z), \qquad z=1+\frac{2t}{\tilde{s}}
\nonumber \\
&=&g_{o}^2(1+(-1)^{s+1}) \frac{2\pi}{\tilde{s}} \int_0^{\tilde{s}} dx 
\frac{\Gamma(s+3-x)}{\Gamma(s+2)\Gamma(2-x)} P_L\left(1-\frac{2x}{\tilde{s}}\right), 
\quad x=-t .
\eea
It can be shown by using the Rodrigues formula:
\bea
P_n(x)=\frac{(-1)^n}{2^n n!}\frac{d^n}{dx^n}(1-x^2)^n,
\eea
that 
\bea
P_L\left(1-\frac{2x}{\tilde{s}}\right) 
=
\frac{(-1)^L}{2^L L!}\left(-\frac{\tilde{s}}{2}\frac{\p}{\p x}\right)^L
Q_L\left(1-\frac{2x}{\tilde{s}}\right) ,
\eea
where $Q_L(x)$ is defined in (\ref{Qfunction}).
Thus we have
\bea
A_L(s)
=g_{o}^2 (1+(-1)^{s+1}) 
\frac{2\pi}{\tilde{s}}  \frac{\tilde{s}^L(-1)^L}{2^{2L} L!}  \int_0^{\tilde{s}} dx 
Q_L\left(1-\frac{2x}{\tilde{s}}\right)
\frac{\p^L}{\p x^L} 
\frac{\Gamma(s+3-x)}{\Gamma(s+2)\Gamma(2-x)} .
\label{preAL}
\eea
The integrand has the same property as we argued to obtain the projection factor in (\ref{exact1}) which is in this case a consequence of the $t \leftrightarrow u$ crossing symmetry originating from the one under an interchange of the final states which are identical with each other.
It enables us to obtain
\bea
A_L(s)
&=& 
\frac{2\pi (1+(-1)^{s+1})(1+(-1)^{L})g_{o}^2\tilde{s}^{L-1}}{2^{2L} L!}   
\nonumber \\
&&\times
\int_0^{\tilde{s}/2} dx Q_L\left(1-\frac{2x}{\tilde{s}}\right)
\left(-\frac{\p}{\p x}\right)^L 
\frac{\Gamma(s+3-x)}{\Gamma(s+2)\Gamma(2-x)} .
\label{AL}
\eea
One immediately notices a similarity to (\ref{exact1}) in the previous section.

As before we take the approximation (\ref{Stirlingapprox}) 
which is valid for large $s$.
The same calculation leads us to  
the cross section:
\bea
\sigma_{open}(s, L)&=&(2L+1)A_L/s
\nonumber \\
&\simeq& (1+(-1)^{s+1})(1+(-1)^{L})
\frac{g_{o}^2 \pi \sqrt{2\pi}e^{-\tilde{s}\ln s/2}}{(\tilde{s}\ln s/2)^{1/2}}
(2L+1)I_{L+1/2}(\tilde{s}\ln s/2) .
\label{partialwavecrosssection4}
\eea

In the argument above we have used the Legendre polynomials 
as a basis of the expansion, which are the four dimensional (part of) spherical harmonics.
Instead one can use $D$-dimensional spherical harmonics which are given by the Gegenbauer polynomials\footnote{
The generating function of the Gegenbauer polynomials is
\bea
(1-2xt+t^2)^{-\lambda}=\sum_{L=0}^\infty C^{\lambda}_L(x) t^L ,
\eea
and the polynomial $C_L^\lambda(x)$ of order $L$ is given explicitly by
\bea
C^{\lambda}_L(x)=
\frac{\Gamma(\lambda+1/2)\Gamma(L+2\lambda)}{\Gamma(2\lambda)\Gamma(L+\lambda+1/2)}
\frac{(-1)^L}{2^LL!}(1-x^2)^{-\lambda+1/2}
\frac{d^L}{dx^L}(1-x^2)^{\lambda-1/2+L},
\eea
which is reduced to the Legendre polynomial for $\lambda=1/2$.
They are normalized by
\bea
\int_{-1}^1 dx (1-x^2)^{\lambda-1/2} [C^{\lambda}_L(x)]^2 
= \frac{2^{1-2\lambda}\pi \Gamma(L+2\lambda)}{(L+\lambda)\Gamma^2(\lambda)\Gamma(L+1)},
\quad \lambda > -1/2 .
\eea
} %
and can repeat the same argument\footnote{
See~\cite{O} for more detailed treatment and discussions.
} to obtain a result generalized to $D$-dimensions. 
In particular, we again have the same projection factor 
as in (\ref{AL}) because the integrand in this case also has 
the same property as in (\ref{preAL})
which follows from the crossing symmetry. 
However instead of doing this we provide an alternative derivation of it in the following.

We start with the Regge limit ($s\gg 1$ with $t$ fixed) of the $s$-channel pole residue of the Veneziano amplitude:
\bea
\text{Im}_{s \sim (N-1)}V(s,t,u) &\sim& (1+(-1)^{s+1}) s^{1+t} . 
\eea
The limit breaks the $t \leftrightarrow u$ symmetry, however we should keep the crossing symmetry which is a reflection of the fact that the final states are identical as we mentioned before. 
We may recover it by introducing the term in the other Regge limit ($s\gg 1$ with $u$ fixed) and have
\bea
\text{Im}_{s \sim (N-1)}V(s,t,u) &\sim&
(1+(-1)^{s+1}) (s^{1+t}+s^{1+u})/2
\nonumber \\
&=&
(1+(-1)^{s+1}) se^{-\frac{\tilde{s}}{2}\ln s} 
(e^{z\frac{\tilde{s}}{2}\ln s}+e^{-z\frac{\tilde{s}}{2}\ln s})/2 .
\label{Reggesym}
\eea
It is easy to verify a formula:
\bea
e^{\xi z}=\sum_{L=0}^{\infty}\frac{2^{3\lambda-1}}{\sqrt{\pi}}\frac{\Gamma(\lambda+1/2)\Gamma(\lambda)^2}{\Gamma(2\lambda)} (L+\lambda) 
\xi^{-\lambda} I_{L+\lambda}(\xi)C_L^\lambda(z),
\eea
where $C_L^\lambda(z)$ is the Gegenbauer polynomial and $\lambda=(D-3)/2$.
We apply this formula together with the one obtained by replacing $z$ with $-z$ (thus we have $(-1)^L$ since $C_L^\lambda(-z)=(-1)^LC_L^\lambda$(z)) to (\ref{Reggesym}).
Then we get a $D$-dimensional generalization of (\ref{partialwavecrosssection4}) that is 
\bea
\sigma(s,L) \propto g_o^2(1+(-1)^{s+1}) (1+(-1)^L)
e^{-\xi}
\xi^{-\lambda} (L+\lambda) I_{L+\lambda}(\xi) , \quad \xi=\frac{\tilde{s}}{2}\ln s .
\label{partialwavecrosssectionD}
\eea
However, we notice here that it is not guaranteed that 
we can always start from an amplitude in the Regge limit. 
In fact, it should be ensured by the fact that 
a function we try to expand in terms of spherical harmonics 
takes a sufficiently small value unless $x$ is small 
as we saw in (\ref{Stirling}).

One can calculate a partial wave amplitude for closed string case in a similar way.
From the Virasoro-Shapiro amplitude: 
\bea
V(s,t,u)=g_{s}^2 \prod_{w=s,t,u}
\frac{\Gamma(-1- w)}{\Gamma(2+ w)}, 
\eea
where $s+t+u=-4$ and $\alp=4$, 
we find 
\bea
\sigma_\text{closed}(s,L) \propto g_{s}^2(1+(-1)^L) s e^{-\tilde{s}\ln s}(\tilde{s}\ln s)^{-\lambda}
(L+\lambda) I_{L+\lambda}(\tilde{s}\ln s) .
\eea
Note the existence of the projection factor originating from the crossing symmetry, but there is no $(1+(-1)^{s+1})$ factor for the lack of the nonplanar diagram.

Among all $L$, the largest contribution is provided by $L=0$ and the cross section asymptotically becomes
\bea
\sigma_\text{open}(s,L=0)
\sim g_o^2 \left(\frac{s}{2}\ln s\right)^{-\lambda-1/2} , \qquad s \gg 1
\label{cross sectionL=0}
\eea
for open string
and
\bea
\sigma_\text{closed}(s,L=0)
\sim
g_{s}^2 s(s\ln s)^{-\lambda-1/2}, \qquad s \gg 1
\eea
for closed string. 
It is evident that the cross section for the open string damps much faster than that of the closed string since the open string has less degrees of freedom than the closed string.

It is also interesting to compare the result in this section 
with that in the previous one.
The cross section with a definite total angular momentum damps more rapidly than the one obtained in the previous section $\sigma(s,J=0) > \sigma_\text{open}(s,L=0)$, since the projection to states 
with a definite total angular momentum is more restricted, and 
the number of states which contribute to the cross section is 
smaller than the case of fixed $J_{12}$. 
As for the $L$-dependence of $\sigma_\text{open}(s,L)$ 
given in (\ref{partialwavecrosssectionD}) we find that it is again basically Gaussian and the same 
as $\sigma(s,J)$ given in (\ref{Jcrosssection}) or (\ref{JGaussian}) when $s$ is large enough
by using the asymptotic form of the modified Bessel function 
given in appendix \ref{Ij}. However, there is a 
slight difference in that $L$ is shifted by $\lambda$. 
It may originate from a quantum effect, but it is interesting 
to clarify its origin.

In the last, we show that we are able to reproduce the formula (\ref{Jcrosssection}) by the partial wave expansion in terms of plane waves which are the eigenfunctions 
corresponding to $J$.
In other words, the partial wave amplitude is obtained as the Fourier transform of the imaginary part of the Veneziano amplitude.
In fact one can explicitly show
\bea
{\mbox{Im}}_{s \sim (N-1)}V(s,t,u) 
= \sum_{J=-\infty}^\infty \cP(s,J) e^{i J \theta},
\eea
where $\cP(s,J)$ is given by (\ref{expression1}) and its non-planar pair, which implies that we have
\bea
\cP(s,J) = \int_0^{2\pi} \frac{d\theta}{2\pi} 
{\mbox{Im}}_{s \sim (N-1)}V(s,t,u) e^{-i J \theta}.
\eea

\section{Summary and discussions}
\label{summary}
\setcounter{equation}{0}

We have calculated the total production cross sections of a single rotating string by both direct and indirect methods.
The former is done with the help of the projection operator to eigenstates of the angular momentum, while the latter is by the optical theorem applied to the four point amplitudes.
Both result in qualitatively 
the same formula which is written in terms of the modified Bessel function whose leading behavior has a Gaussian profile with respect to the angular momentum giving the form factor of a string corresponding to an object of size $\sqrt{\ln s}$. 
The cross sections with a fixed angular momentum damp as center of mass energy $\sqrt{s}$ grows, in contrast to the inclusive case in which the cross section stays constant or raises linearly in $s$ 
in open and closed string case, respectively. 
In our case two methods provide similar results as expected 
and the latter is easier than the former. However, we emphasize here 
that the former has an advantage that it can be generalized 
to cases where the optical theorem cannot be applied in a 
straightforward manner, like specifying other quantum numbers 
than the angular momentum, or a mutli-string production cross section corresponding to a loop diagram which is in fact treated 
in \cite{Manes:2003mw}, \cite{Manes:2004nd}.

It would be intriguing to compare our result with the production 
cross section of a black hole with an angular momentum $L$ 
which is evaluated geometrically. 
It is known that when $L$ is not specified, 
the production cross section of a string 
and that of a black hole have different $s$-dependence 
even at the correspondence point \cite{Dimopoulos:2001qe}. 
Now our result enables us to compare 
$L$-dependence of these two as well. 
Indeed, they have again quite different behavior 
as functions of $L$, because 
the production cross section of a 
string is essentially 
Gaussian, while that of a black hole raises with $L$ 
at least in small $L$ region. Here it should be stressed 
that we do not take account of string loop effects. 
In fact, in high energy regime higher genus amplitudes 
are known to be dominant and hence it is evident that 
in order to compare the results of string and black hole 
at the correspondence point, we have to control 
higher genus amplitudes in a systematic way. 
Therefore it would be quite interesting to examine 
how $L$-dependence of the production cross section 
of a
string would change drastically 
via higher genus amplitudes by using, {\it e.g.} the eikonal amplitude 
developed in \cite{Amati:1987wq}, \cite{Amati:1987uf}, 
assuming that the correspondence works.

Our result may have lots of potential applications.
High energy string scattering amplitudes have been basic ingredients in arguments of the connection between string theory and hadron physics through the AdS/CFT correspondence \cite{Polchinski:2001tt}, \cite{Polchinski:2002jw}, \cite{Brower:2006ea}, \cite{Giddings:2002cd}. 
To apply not only to these hadron physics but also to cosmic string and Polymer physics, etc are also interesting.

Although we do not consider their effects, if scattering energy is high enough to produce some non-perturbative objects like D-branes and NS-branes, then we have to take account of these effects.
It would be interesting to investigate these effects on the cross sections.
Following \cite{Manes:2003mw}, \cite{Manes:2004nd} it is also interesting to find the form factor or a profile of the wave function of a rotating string.
The superstring generalization seems valuable for future use, though essential feature may be captured by bosonic string theory.
 
\section*{Acknowledgements}
The authors would like to thank Kin-ya Oda for collaboration in an early stage of the present work.
We are also grateful to Hikaru Kawai, Tamiaki Yoneya, and Keiji Igi 
for valuable discussions.
One of the authors T.M would like to thank Takayuki Hirayama, Feng-Li Lin, Dan Tomino for discussions and comments.
He also thanks to Hiroshi Itoyama for encouragement. 
The authors thank the Yukawa Institute for Theoretical Physics at Kyoto University. Discussions during the YITP workshop YITP-W-07-05 on ``String Theory and Quantum Field Theory'' were useful to complete this work.
The work of T.M is supported by the Taiwan's National Science Council under grants NSC95-2811-M-003-005.

\appendix

\section{Trace calculation}
\label{tracecalc}
\setcounter{equation}{0}

We would like to evaluate (non-zero modes)
\bea
\tr[V(k,\rho)z^{\hat{J}}V(p,v)w^{\hat{N}}]
=\prod_{n=1}\left[T_n^{1-2} \prod_{i=0, 3, \cdots} T_n^i \right],
\label{trace}
\eea
where $V$ is a vertex operator for a string state.

The normal ordered tachyon vertex operator is 
\bea
V(k,\rho)=
\exp\left(
\sqrt{2}\sum_{n=1}^{\infty}\frac{1}{n}k \cdot \alpha_{-n} \rho^n
\right)
\exp\left(
-\sqrt{2}\sum_{n=1}^{\infty}\frac{1}{n}k \cdot \alpha_{n} \rho^{-n}
\right) .
\eea
Note here we take $\alp=1$ convention as in the main part of the present paper and set $g_o=1$ for simplicity.
We first focus on the 1,2 components and use an arrow to represent the $(1,2)$ component of vectors such as ${\vec p}=(p_1,p_2)$, etc. 
The relevant part is
\bea
T_n^{1-2}(k,p;\rho,z,v,w)&=&\tr\left[
\exp\left(\sqrt{2}\frac{\rho^n}{n} \vec{k} \cdot \vec{\alpha}_{-n}\right)
\exp\left(-\sqrt{2}\frac{\rho^{-n}}{n} \vec{k} \cdot \vec{\alpha}_{n}\right)
\exp\left(-\frac{i \ln z}{n} \vec{\alpha}_{-n} \times \vec{\alpha}_{n}\right)
\right.
\nonumber \\
&&\times
\left.
\exp\left(\sqrt{2}\frac{v^n}{n} \vec{p} \cdot \vec{\alpha}_{-n}\right)
\exp\left(-\sqrt{2}\frac{v^{-n}}{n} \vec{p} \cdot \vec{\alpha}_{n}\right)
\exp\left(\ln w \vec{\alpha}_{-n} \cdot \vec{\alpha}_{n}\right)
\right] .
\eea
To calculate the trace it is convenient to use the coherent state basis\footnote{
In the worldsheet point of view, (\ref{trace}) is a finite temperature Green function (two point function on cylinder) between the rotated vertex $z^{-\hat{J}}V(k,\rho)z^{\hat{J}}$ and  $V(p,v)$ 
with a worldsheet Hamiltonian $\beta H=-\ln w \hat{N}-\ln z \hat{J}$  \cite{Russo:1994ev}. Therefore one can use, as an alternative way, 
the Wick's theorem in terms of the finite temperature Green function 
to find (\ref{T12}). 
}: %
\bea
\tr(A)=\int \frac{d^2z_1}{\pi} \int \frac{d^2z_2}{\pi} e^{-|z_1|^2-|z_2|^2} \la z_1z_2|A|z_1z_2\ra,
\eea
where
\bea
|z_1z_2\ra = 
\exp\left(\frac{1}{\sqrt{n}}\vec{z} \cdot \vec{\alpha}_{-n}\right)|0\ra.
\eea
Using the Baker-Campbell-Hausdorff (BCH) formula it yields
\bea
T_n^{1-2}
&=&\int \frac{d^2z_1}{\pi} \int \frac{d^2z_2}{\pi} e^{-|z_1|^2-|z_2|^2}
\exp\left(\sqrt{2}\frac{\rho^n}{\sqrt{n}} \vec{k} \cdot \vec{z}^*
-\sqrt{2}\frac{w^nv^{-n}}{\sqrt{n}} \vec{p} \cdot \vec{z} \right)
\nonumber \\
&&\times
\la0|
\exp\left(\left(-\sqrt{2}\frac{\rho^{-n}}{n} \vec{k}
+\frac{\vec{z}^*}{\sqrt{n}}\right)\cdot \vec{\alpha}_{n}\right)
\exp\left(-\frac{i \ln z}{n} \vec{\alpha}_{-n} \times \vec{\alpha}_{n}\right)
\nonumber \\
&&\times
\exp\left(\left(\frac{w^n}{\sqrt{n}}\vec{z}
+\sqrt{2}\frac{v^n}{n} \vec{p} \right) \cdot \vec{\alpha}_{-n}\right)
|0\ra.
\eea
To compute this we shall use the formula which is easy to verify
\bea
&&
\exp\left(-\frac{iA}{n}\vec{\alpha}_{-n}\times\vec{\alpha}_n\right)
\exp\left(\vec{B}\cdot \vec{\alpha}_{-n}\right)|0\ra
\nonumber \\
&&
=
\exp\left(
\left(\cos(iA)\alpha_{-n}^1 +\sin(iA)\alpha_{-n}^2 \right)B_1
+\left(-\sin(iA)\alpha_{-n}^1+\cos(iA)\alpha_{-n}^2 \right)B_2
\right)
|0\ra,
\eea
and again we use the BCH formula to eliminate all the creation and annihilation operators.
Then using the formula: 
\bea
&&\int \frac{d^2z_1}{\pi} \int \frac{d^2z_2}{\pi} 
e^{-c_1|z_1|^2+a_1z_1+b_1z_1^*-c_2|z_2|^2+a_2z_2+b_2z_2^*+dz_1z_2^*+ez_1^*z_2} 
\nonumber \\
&&=\frac{1}{c_1c_2-ed}
\exp\left(\frac{a_2b_2c_1+a_1b_1c_2+a_1b_2e+a_2b_1d}{c_1c_2-ed}\right),
\eea
we arrive at
\bea
T_n^{1-2}&=&\frac{1}{1-2cw^n+w^{2n}}
\exp\left[
-\frac{2}{n(1-2cw^n+w^{2n})}
\left\{
s{\vec p}\times{\vec k}(v/\rho)^n-sw^{2n}{\vec p}\times{\vec k}(\rho/v)^n
\right. \right.
\nonumber \\
&&\left.\left.
+(c-w^n){\vec p} \cdot {\vec k} (v/\rho)^n 
+w^n(1-cw^n){\vec p} \cdot {\vec k}(\rho/v)^n
+w^n(c-w^n)({\vec k} \cdot {\vec k}+{\vec p} \cdot {\vec p})
\right\}
\right] ,\nonumber\\
\label{T12}
\eea
where 
\bea
c:
=\cosh(\ln(z))=\frac{1}{2}(z+z^{-1}) , 
\qquad
s:
=i\sinh(\ln(z))=\frac{i}{2}(z-z^{-1}) .
\eea
Taking product with respect to $n$ by using a formula (which is verified by expanding both sides):
\bea
\sum_{n=1}^\infty \frac{V^n}{n(1-zw^n)(1-z^{-1}w^n)}
=
-\sum_{m=0}^\infty \frac{z^{m+1}-z^{-m-1}}{z-z^{-1}}\ln (1-Vw^m),
\eea
one obtains 
\bea
\prod_{n=1}^\infty T^{1-2}_n(\vec{k},\vec{p};\rho,z,v,w)=
\tilde{f}(z,w)^{-1}
[\hat{\Psi}(z,v/\rho,w)]^{\vec{p} \cdot \vec{k}}
[\hat{\Phi}(z,v/\rho,w)]^{i \vec{p} \times \vec{k}},
\eea
where 
\bea
\tilde{f}(z,w) := \prod_{m=1}^\infty(1-zw^m)(1-z^{-1}w^m) ,
\label{tildef}
\eea
\bea
\hat{\Psi}(z,v,w) := (1-v)^{z+z^{-1}} \prod_{m=1}^\infty
\frac{
(1-vw^m)^{z^{m+1}+z^{-m-1}}
(1-w^m/v)^{z^{m-1}+z^{-m+1}}
}{
(1-w^m)^{2(z^m+z^{-m})}
} ,
\label{Psihat}
\eea
and 
\bea
\hat{\Phi}(z,v,w) := (1-v)^{z-z^{-1}}
\prod_{m=1}^\infty
(1-vw^m)^{z^{m+1}-z^{-m-1}}
(1-w^m/v)^{-(z^{m-1}-z^{-m+1})} .
\eea
Here we have used the on-shell condition $\vec{p}+\vec{k}=0$ to convert $\vec{p}^2+\vec{k}^2=-2 \vec{p} \cdot \vec{k}$.

It is easy to obtain contributions from other components $i=0, 3,\ldots,d$  by setting $z=1$ in the above expressions, then
\bea
T_n^i=\frac{1}{1-w^n}
\exp\left[
-\frac{2}{n(1-w^n)}\left\{p^i  k^i (v/\rho)^n 
+w^n p^i  k^i(\rho/v)^n
+w^n(k^i  k^i+p^i  p^i)
\right\}
\right] .
\eea
One may get after taking product with respect to $n$ 
\bea
\prod_{n=1}^\infty T^i_n(k_i,p_i;\rho,1,v,w)=[f(w)]^{-1}[\hat{\psi}(v/\rho,w)]^{2 p_ik_i},
\eea
for each $i=0,3,\ldots,d$ and again we have used the on-shell condition $p_i+k_i=0$.
Here the Dedekind function $f(w)$ and $\hat{\psi}(v,w)$ are given as
\bea
f(w):= \prod_{n=1}^\infty(1-w^n),
\label{Dedekind}
\eea
and
\bea
\hat{\psi}(v,w):= (1-v)\prod_{m=1}^\infty
\frac{(1-vw^m)(1-w^m/v)}{(1-w^m)^2} .
\label{psihat}
\eea

Combining all we obtain
\bea
\tr[V(k,\rho)z^{\hat{J}}V(p,v)w^{\hat{N}}]
&=&
[f(w)]^{-d+1}\tilde{f}(z,w)^{-1}
\nonumber \\
&&\times
\prod_{i=0,3,\ldots,d}[\hat{\psi}(v/\rho,w)]^{2p_ik_i}
[\hat{\Psi}(z,v/\rho,w)]^{\vec{p} \cdot \vec{k}}
[\hat{\Phi}(z,v/\rho,w)]^{i \vec{p} \times \vec{k}} .\nonumber\\
\eea

\section {Asymptotic form of the modified Bessel function}
\label{Ij}
\setcounter{equation}{0}
The modified Bessel function has a following integral representation for integer $\nu$:
\bea
I_\nu(z)
=\text{Re}\left[
\frac{1}{\pi}\int_0^\pi d\theta e^{z\cos \theta+i \nu \theta}
\right].
\eea
Let us derive an asymptotic formula for large $z$ by a saddle point method.
Let $ f(\theta) := z\cos \theta +i \nu \theta$, we find a saddle point $\theta_0$ which is determined by 
\bea
\theta_0=\sin^{-1}(i\nu/z)=i\sinh^{-1}(\nu/z)
=i\ln\left(\frac{\nu}{z}+\sqrt{1+\frac{\nu^2}{z^2}}\right),
\eea
which is pure imaginary for real $\nu, z$.
This can be an end point of the integration region by an appropriate modification of the contour.
The half of the saddle point thus contributes to the solution.
Let $\theta_0$ be a saddle point.
The asymptotic form is given as 
\bea
I_\nu(z) \simeq 
\frac{1}{2}Re\left[
\frac{e^{f(\theta_0)}}{\pi}
\sqrt{
\frac{2\pi}{|f''(\theta_0)|}}
\right],
\eea
where
\bea
\text{Re}\exp(f(\theta_0))&=&e^{z\sqrt{1+\frac{\nu^2}{z^2}}-\nu \sinh^{-1}(\nu/z)}
\nonumber \\
&=&e^{z\sqrt{1+\frac{\nu^2}{z^2}}-\nu \ln\left(\frac{\nu}{z}+\sqrt{1+\frac{\nu^2}{z^2}}\right)},
\eea
and
\bea
f''(\theta_0)=-z\cos(\theta_0)=-z\sqrt{1+\frac{\nu^2}{z^2}},
\eea
thus
\bea
I_\nu(z) \sim 
\frac{1}
{\sqrt{2\pi z\sqrt{1+\frac{\nu^2}{z^2}}}}
\exp\left[z\sqrt{1+\frac{\nu^2}{z^2}}-\nu \ln\left(\frac{\nu}{z}+\sqrt{1+\frac{\nu^2}{z^2}}\right)\right] .
\label{asymptoticMBessel}
\eea


 \end{document}